\documentclass{mem}
\usepackage[]{fontenc}
\usepackage[latin1]{inputenc}
\usepackage{amsmath}
\usepackage{graphicx}
\usepackage{amssymb}
\usepackage[authoryear]{natbib}

\makeatletter

\providecommand{\tabularnewline}{\\}

\usepackage{natbib}
\usepackage{graphicx}
\usepackage[a4paper]{hyperref}

\makeatother
\begin{document}
\title{Synthetic Super AGB Stars}

\author{Robert G. Izzard$^{1,2}$ and Arend Jan T. Poelarends$^1$}

\offprints{\tt{R.G.Izzard@phys.uu.nl}}

\institute{1. Sterrenkundig~Instituut, Universiteit~Utrecht, P.O.~Box~80000, 3508~TA~Utrecht, The~Netherlands.\\
2. The Carolune Institute for Quality Astronomy, \tt{www.ciqua.org}}

\abstract{We describe our first attempt at modelling nucleosynthesis
in massive AGB stars which have undergone core carbon burning, the
super-AGB stars. We fit a synthetic model to detailed stellar evolution
models in the mass range $9\leq M/\mathrm{M_{\odot}}\leq11.5$ ($Z=0.02$),
and extrapolate these fits to the end of the AGB. We determine the
number of thermal pulses and AGB lifetime as a function of mass and
mass-loss prescription. Our preliminary nucleosynthesis calculations
show that, for a reasonable mass-loss rate, the effect of hot-bottom
burning in super-AGB stars on the integrated yield of a stellar population
is not large. There are many uncertainties, such as mass-loss and
convective overshooting, which prevent accurate yield calculations.
However, as potential progenitors of electron-capture supernovae,
these stars may contribute $7\%$ of non-type-Ia supernovae.

}

\authorrunning{R.G.~Izzard and A.J.T.~Poelarends}    

\titlerunning{Synthetic SAGB Stars}

\maketitle

\section{Introduction}

Stars are traditionally divided into those which explode and those
which do not. Stars which explode as supernovae by a core-collapse
mechanism are the massive stars, while their lower-mass cousins enter
a thermally pulsing asymptotic giant branch (TPAGB) phase where rapid
mass-loss ends their evolution. It is difficult to specify the exact
mass boundary between these populations, with estimates ranging from
$7\mathrm{\, M_{\odot}}$ \citep{2000A&AS..141..371G} up to more
than $11\mathrm{\, M_{\odot}}$ \citep{1999ApJ...515..381R}, depending
on metallicity and convective overshooting. 

The problem relates to the fate of the star after carbon ignition,
which depends on the degeneracy of the core. In high-mass stars the
non-degenerate core burns carbon, then neon, oxygen and silicon. A
core-collapse supernova soon follows, leaving a neutron star or black
hole. A lower-mass, but massive enough to ignite carbon, (partially)
degenerate core burns carbon in a series of flashes (e.g. \citealp{2006A&A...448..717S})
and then moves to a \emph{super} thermally pulsing AGB (SAGB) stage,
with double-shell burning above a degenerate oxygen-neon core, where
mass-loss terminates the evolution, leaving a white dwarf (Siess,
this volume). In a small number of stars the oxygen-neon core may
grow beyond $1.368\mathrm{\, M_{\odot}}$ during the pulsing phase,
which leads to a collapse of the core due to electron capture on $^{24}\textrm{Mg}$
(Poelarends, this volume). Detailed models of these stars have been
constructed by \citet{1994ApJ...434..306G,1996ApJ...460..489R,1997ApJ...485..765G,1997ApJ...489..772I}
and more recently by \citet{2004MmSAI..75..694E} who consider SAGB
stars as observable supernova progenitors (see also Poelarends, this
volume).

Once carbon ignition and second dredge-up have finished, SAGB stars
pulse in much the same way as normal TPAGB stars, albeit with a shorter
interpulse period. By virtue of their high mass, one expects high
temperatures at the base of the convective envelope and associated
hot-bottom burning (HBB; e.g. \citealp{1995ApJ...442L..21B}). Their
envelopes should be processed by the hydrogen-burning CNO, NeNa and
MgAl cycles. For studies of Galactic chemical evolution, these stars
may be an important source of nitrogen, sodium and aluminium and they
may play a part in the globular cluster $\textrm{Na}-\textrm{O}$
anticorrelation mystery (D'Antona, this volume; \citealp{2005ApJ...635L.149V}).
It is unclear whether third dredge-up occurs in SAGB stars -- current
models suggest either it does (Doherty, this volume), does not (our
models and Siess, this volume), or is inefficient \citep{1996ApJ...460..489R}.
There are no detailed studies of chemical yields from SAGB stars,
probably because detailed stellar models take a long time and are
difficult to construct, and suffer from the usual uncertainty due
to mass-loss and convective overshooting. A synthetic modelling technique
speeds up modelling and enables us to explore the uncertain parameter
space.

In this paper we calculate the chemical yields of SAGB stars using
a synthetic model based on the AGB model of \citet[I04]{Izzard_et_al_2003b_AGBs}.
We approximate stellar structural variables with formulae and interpolation
tables, and use a simple model for HBB to follow the CNO, NeNa and
MgAl cycles and surface abundances of C, N, O, Ne, Na, Mg, Al and
Si. The synthetic model is then used to extrapolate evolution beyond
the detailed models to the end of the SAGB phase. This is possible
because the structure of AGB stars is such that after a number of
pulses, the evolution reaches a limit cycle \citep{1996ApJ...460..489R}.
We calculate supernova rates and chemical yields, and also change
the input physics (especially the mass-loss rate) to determine the
effect of uncertainties. Such a parameter space exploration is currently
impossible with a normal stellar evolution code, because the CPU time
required is simply too large.

\section{Models, full and synthetic}

Our full evolution models were constructed with the STERN code \citep*{2000ApJ...528..368H}.
We constructed models of mass $8.5$, $9.0$, $10.0$ and $11.5\mathrm{\, M_{\odot}}$
which undergo $12$,$26$,$10$,$16$ pulses respectively, with metallicity
$Z=0.02$, no convective overshooting, no mass loss and no rotation.
Further description of these models can be found in Poelarends et
al. (this volume).

Our synthetic models are based on those of I04 with some updates and
changes for the SAGB phase (a detailed description will be found in
a later paper; Izzard and Poelarends, in preparation). The luminosity
formula was altered to fit our detailed models, and the radius follows
from $\log R\sim\log L$. The initial and post-second-dredge-up abundances,
core mass at the end of core helium burning and core mass at the first
thermal pulse, are interpolated from tables based on the detailed
models. Stars with a helium core mass above $1.6\mathrm{\, M_{\odot}}$
during the early AGB ignite carbon, while stars with a degenerate
oxygen-neon core with mass greater than $1.38\mathrm{\, M_{\odot}}$
collapse to neutron stars. We assume there is no third dredge up. 

The synthetic HBB model was described in I04 but we use the latest
version which includes the NeNa and MgAl cycles as well as CNO. It
approximates the burn-mix-burn-mix\ldots{} cycle in a real convective
envelope (which has a thin HBB shell at the base) as a single burn-mix
event during each interpulse period, with a large fraction of the
envelope burned for a given time. The temperature and density at the
base of the envelope are fitted to simple formulae of the form $1-\exp\left(-N_{\textrm{TP}}\right)$
which approaches a constant as $N_{\textrm{TP}}$, the number of thermal
pulses, increases. The fraction and the burn time are calibrated as
a function of stellar mass, as in I04.

We apply one of the following mass-loss prescriptions during the SAGB
phase: none, original \citet[VW93]{1993ApJ...413..641V}, VW93 but
\citet[K02]{Parameterising_3DUP_Karakas_Lattanzio_Pols} variant,
Reimers with $\eta=1$ or $\eta=5$ \citep{1975psae.book..229R} or
\citet{1991A&A...244L..43B} with $\eta=0.1$. Prior to the SAGB,
we either apply no mass loss, or the compilation of \citet*[H02]{2002MNRAS_329_897H}.
In the cases where mass loss does not expose the core before it grows
to $1.368\mathrm{\, M_{\odot}}$, the core is quietly converted into
a neutron star while the envelope is ejected to space (this is the
subject of some debate e.g. \citealp{1987ApJ...322..206N}, \citealp*{2005A&A...435..231G}).

\section{Results}

\begin{figure}
\includegraphics[bb=50bp 120bp 554bp 770bp,scale=0.32,angle=270]{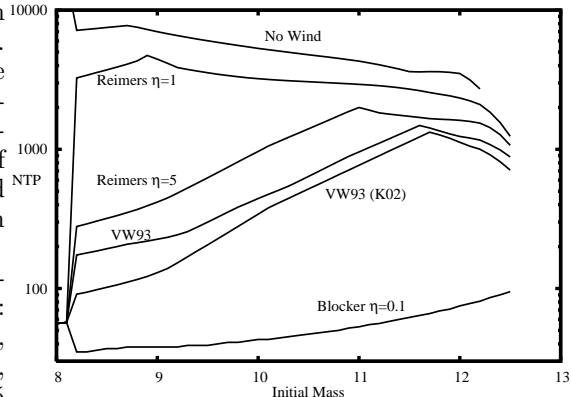}

\caption{\label{cap:NTP}Number of thermal pulses as a function of initial
stellar mass for different mass-loss prescriptions: no wind at all
or pre-SAGB wind of \citet{2002MNRAS_329_897H} with Reimers ($\eta=1$
or $5$), VW93 or Bl\"ocker mass-loss during the SAGB. }
\end{figure}

We are confident that our extrapolation of the stellar structure (luminosity,
radius, mass and core mass evolution) is reasonable, within the uncertainty
that is mass loss, because a very similar model works well for lower-mass
AGB stars (I04). The mass-loss rate for SAGB stars is unknown, so
we consider all the possibilities. Figure \ref{cap:NTP} shows the
number of thermal pulses during the SAGB phase which ranges from $30$
(Bl\"ocker, with SAGB lifetimes of $4,000$ to $5\times10^{4}$ years)
to $4,000$ (Reimers $\eta=1$, with lifetimes of $10^{5}$ to $10^{6}$
years), or $8,000$ with no mass loss.%
 Mass loss during core helium burning, prior to the SAGB, is included
in the H02 prescription and is not negligible for stars above $8\mathrm{\, M_{\odot}}$
-- they lose around $0.5\mathrm{\, M_{\odot}}$ during this phase,
which affects the subsequent evolution and HBB.

One way of constraining the value of mass-loss rate is through supernova
counts. If there is little mass loss during the SAGB, many stars'
cores should reach the electron capture limit of $1.368\mathrm{\, M_{\odot}}$
before their envelope is lost. In the limit of no mass loss during
the SAGB, we find the ratio of the electron-capture to type-II supernova
rates is about one%
\footnote{We do not include an upper mass limit for SNe, so ratios quoted are
lower limits.%
}. With the VW93 (K02 variant) mass-loss the ratio drops to $7\%$.
If electron capture supernovae are distinguishable from normal core-collapse
types (and accretion induced collapses in binaries; \citealp{2004ApJ...612.1044P})
then we can constrain the mass-loss rate%
\footnote{We only model single stars here, binary interactions will alter the
result. %
}.

\begin{figure*}
\begin{tabular}{cc}
\includegraphics[bb=50bp 160bp 554bp 770bp,scale=0.3,angle=270]{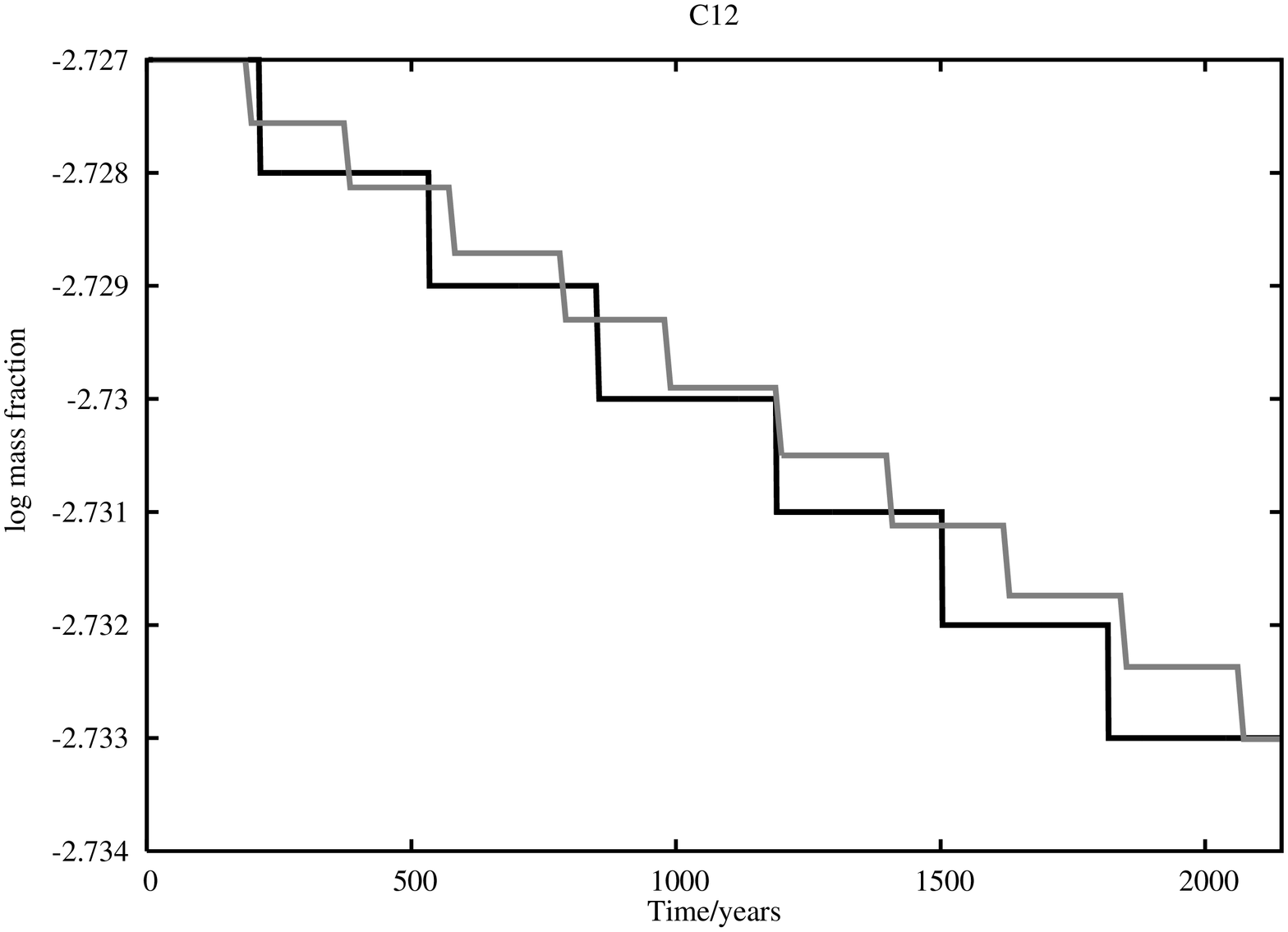}&
\includegraphics[bb=50bp 130bp 554bp 770bp,scale=0.3,angle=270]{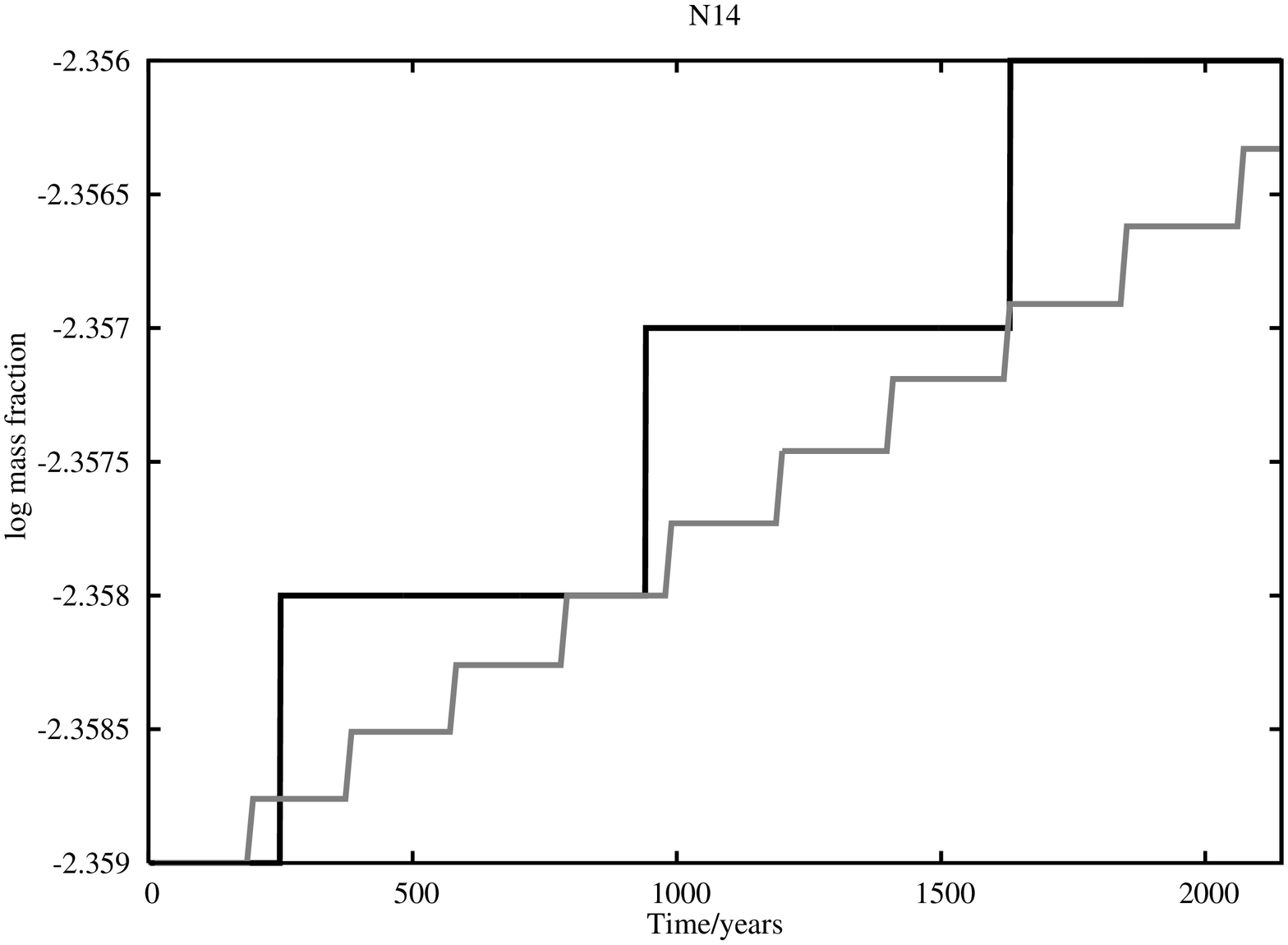}\tabularnewline
\includegraphics[bb=50bp 160bp 554bp 770bp,scale=0.3,angle=270]{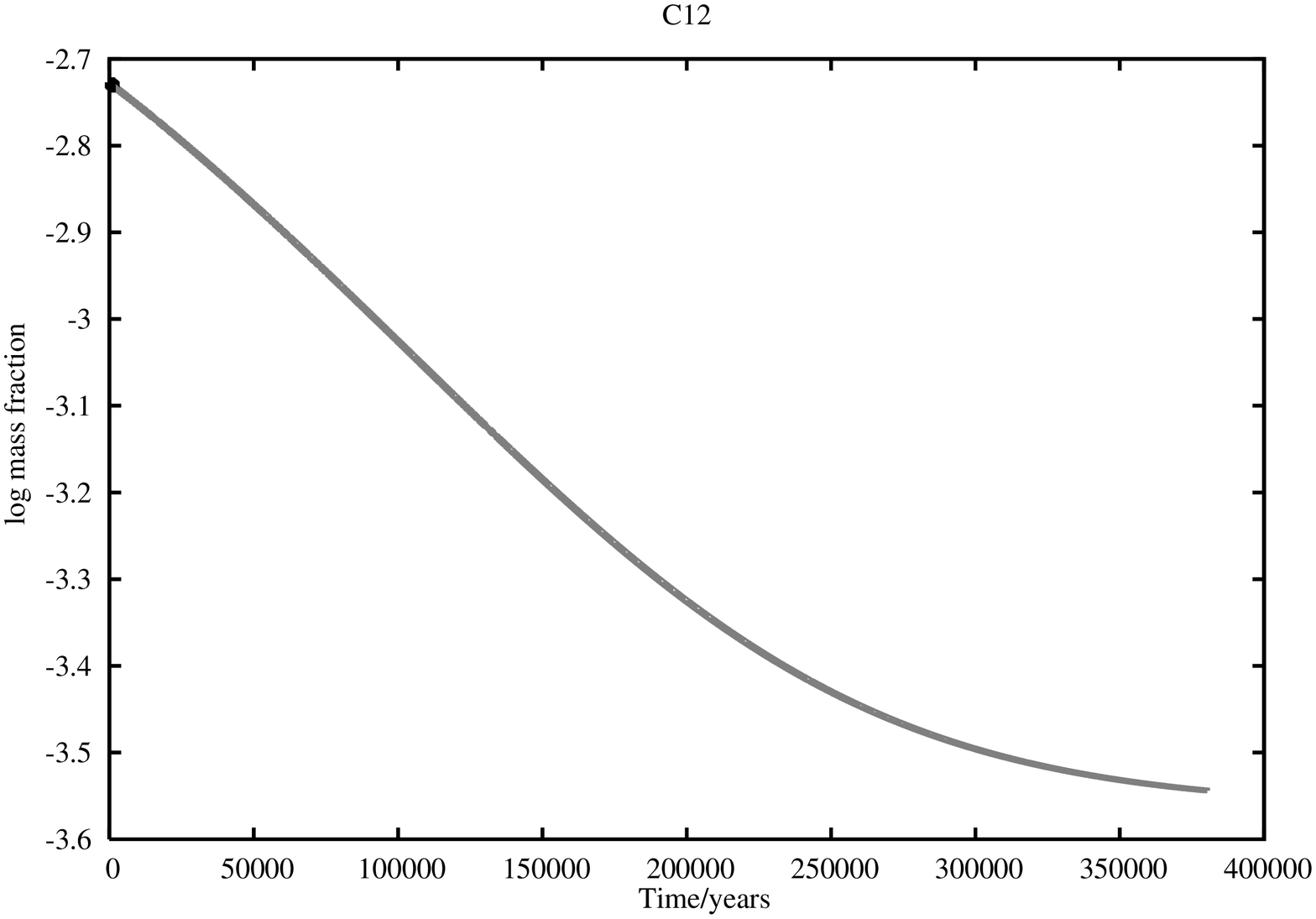}&
\includegraphics[bb=50bp 130bp 554bp 770bp,scale=0.3,angle=270]{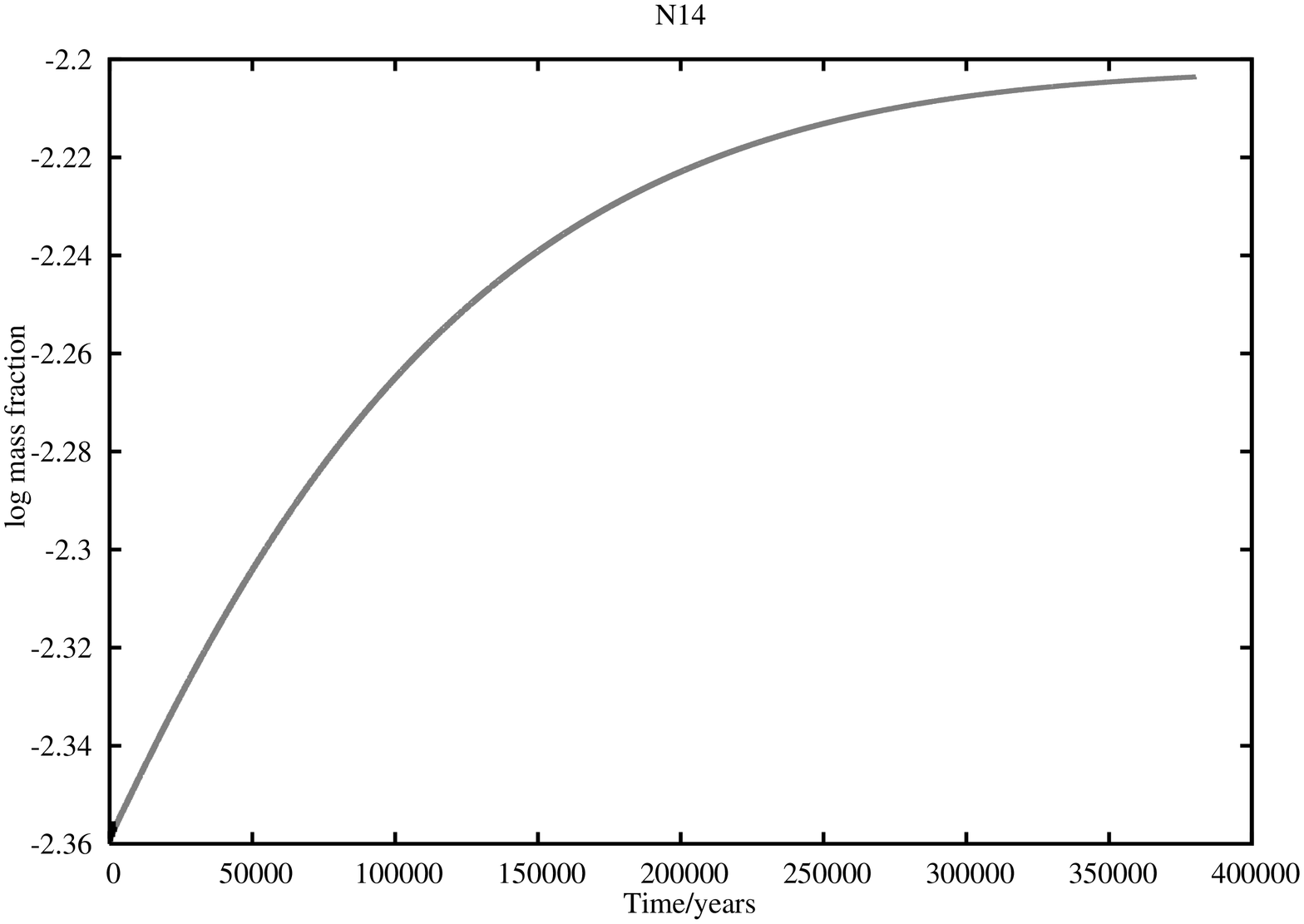}\tabularnewline
\end{tabular}

\caption{\label{cap:Extapolation}Log surface abundance by mass fraction vs
time for $^{12}\textrm{C}$ and $^{14}\textrm{N}$ for our $10\mathrm{\, M_{\odot}}$
detailed (black lines) and synthetic (grey lines) models, both with
no mass loss. The top panels show the result of our HBB calibration,
the bottom panels the same models extrapolated to the end of the SAGB.}
\end{figure*}

In figure \ref{cap:Extapolation} we show the results of our HBB calibration
and extrapolation for a $10\mathrm{\, M_{\odot}}$ star. The detailed
models truncate their abundance output to the nearest $10^{-3}$ in
the log, which causes the stepping behaviour, while the stepping in
the synthetic model is because HBB is done at the end of each pulse.
Ten pulses is just enough to calibrate the isotopes which change rapidly,
$^{13}\textrm{C}$, $^{17}\textrm{O}$ and $^{21}\textrm{Ne}$ are
also useful in this regard, while other isotopes such as $^{16}\textrm{O}$
are not. We have to assume the other species ($^{20,22}\textrm{Ne}$,
$^{23}\textrm{Na}$, $\textrm{Mg}$ and $\textrm{Al}$) follow from
this calibration. The situation is little better in the $9\mathrm{\, M_{\odot}}$
star: it has $26$ pulses, but its lower temperature slows the burning.%

\begin{figure*}
\begin{tabular}{cc}
\includegraphics[bb=50bp 160bp 554bp 770bp,scale=0.3,angle=270]{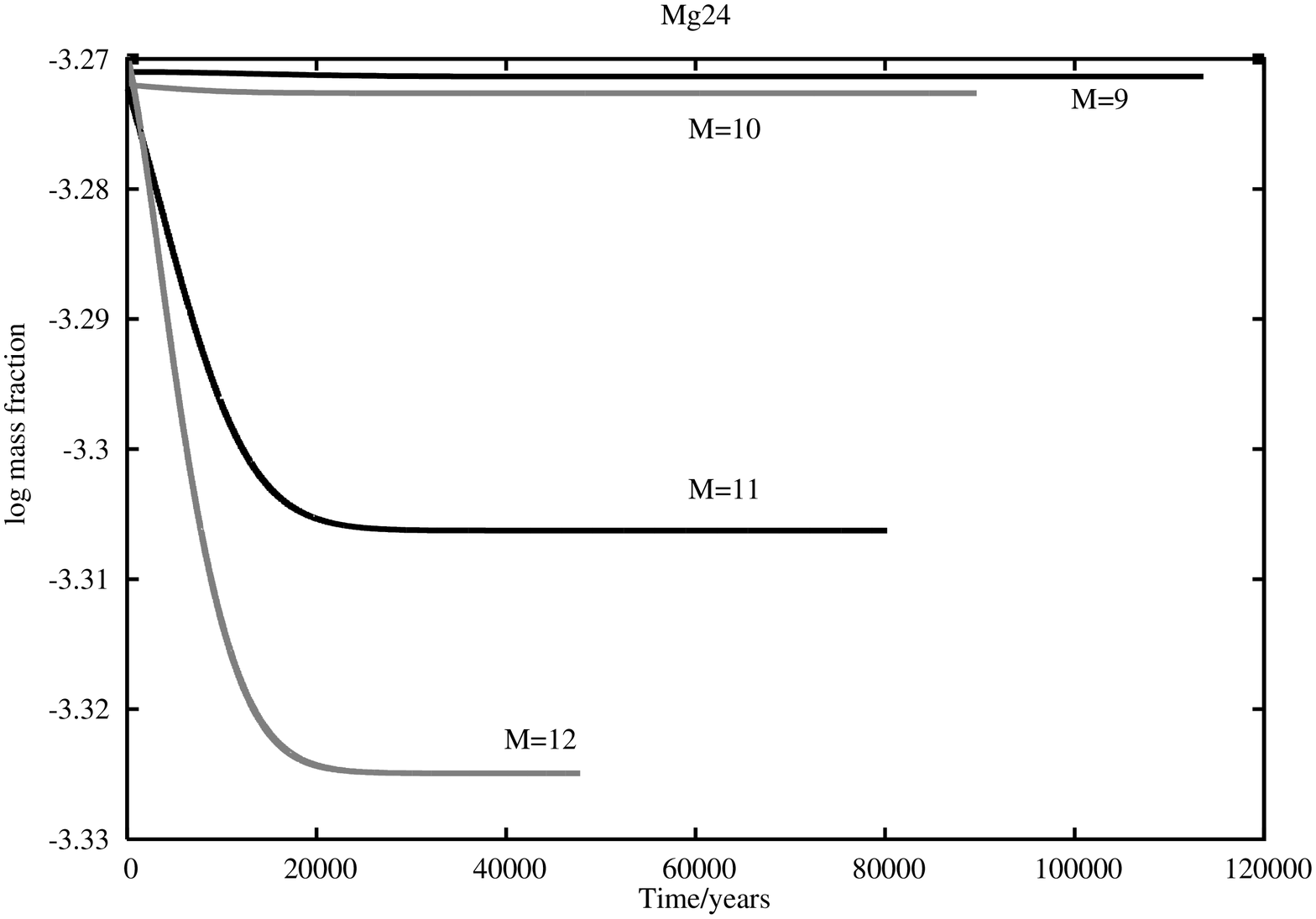}&
\includegraphics[bb=50bp 130bp 554bp 770bp,scale=0.3,angle=270]{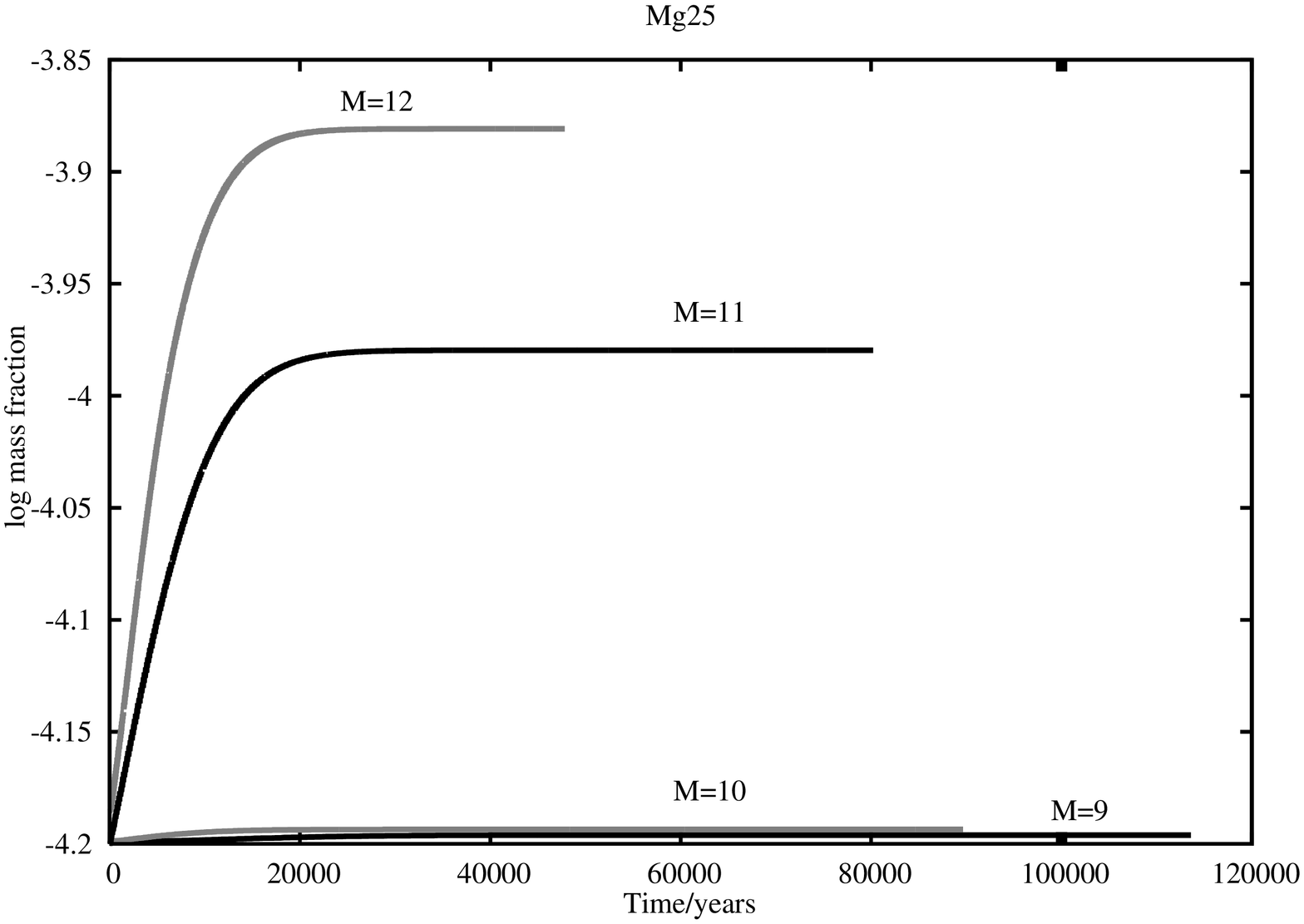}\tabularnewline
\end{tabular}

\caption{\label{cap:magnesium}The surface abundance of $^{24}\textrm{Mg}$
and $^{25}\textrm{Mg}$ during the super-AGB phase for the initial
mass range $9\leq M/\mathrm{M_{\odot}}\leq12$. Hot-bottom burning
converts $^{24}\textrm{Mg}$ into $^{25}\textrm{Mg}$, while mass-loss
stops the burning and terminates the evolution.}
\end{figure*}

Finally, we consider the magnesium isotopes. Figure \ref{cap:magnesium}
shows the surface abundance of $^{24}\textrm{Mg}$ and $^{25}\textrm{Mg}$
as a function of time for $9\leq M/\mathrm{M_{\odot}}\leq12$ with
H02 mass loss prior to the SAGB, and VW93 (K02 variant) mass loss
during the SAGB. Only the $11$ and $12\mathrm{\, M_{\odot}}$ models
show significant burning of $^{24}\textrm{Mg}$ to $^{25}\textrm{Mg}$,
with up to a factor of two increase. For $M\leq10\mathrm{\, M_{\odot}}$
the HBB is simply not hot enough to enable the MgAl cycle. At all
masses, rapid mass loss turns off the MgAl cycle after about $2\times10^{4}\,\textrm{years}$%
\footnote{As in I04 we modulate the temperature with a factor $\left(M_{\textrm{env}}/M_{\textrm{env,1TP}}\right)^{0.02}$
to turn off HBB as mass is lost.%
}. 

\section{Discussion}

This was our first attempt to model these stars synthetically and,
at least for the structure variables such as luminosity, radius, core
mass (and growth), we have confidence in our model. However our HBB
model is not as solid, given that we have extrapolated forward by
a factor of ten or more in time compared to our detailed models. It
is clear that we must extend our detailed models to at least cover,
say, half the evolution (which may still be many thousand pulses),
and see if the synthetic model predictions match the detailed model%
\footnote{The numerical problems seen in figure \ref{cap:Extapolation} will
become irrelevant in that case.%
}. 

Even when we have extended the detailed models, we will still suffer
from the mass-loss uncertainty. It seems, from figure \ref{cap:NTP},
we can get as many pulses as we like. New observations are providing
insight into the mass-loss rates in oxygen-rich AGB stars and we integrate
these new rates into our next calculations \citep{2005A&A...438..273V}.
We have also neglected the problem of convective overshooting, which
reduces the mass for formation of SAGB stars by about $2\mathrm{\, M_{\odot}}$
(Siess, this volume). The initial mass function is a steep power law
in mass, so a lower mass limit means more SAGB stars, and their lifetimes
will be longer (lower mass means smaller luminosity, radius and $\dot{M}$).

Our preliminary results suggest that SAGB stars are not very important
either for Galactic or globular cluster chemical evolution. The bulk
of element production comes from either lower-mass AGB stars or higher-mass
stars and type II supernovae. This conclusion may change if SAGB stars
suffer either third dredge-up, or dredge-out, a phenomenon where the
carbon flash causes helium-burned material to be mixed to the surface
\citep{1999ApJ...515..381R,2006A&A...448..717S}. Also, we only consider
envelope ejection for a collapsing oxygen-neon core. In reality, some
of the core may be ejected too, particularly if some carbon remains
after the core flashes \citep*{2005A&A...435..231G}.

\section{Conclusions}

In the context of galactic chemical evolution, chemical yields from
hot-bottom burning SAGB stars are not important, at least for most
isotopes. If mass-loss rates are much lower than those expected from
extrapolation of normal AGB rates, if there is dredge-up or dredge-out,
or if our simple extrapolations fail, this conclusion may be premature.
We are working to extend our detailed model set to remove the extrapolation
problem.

The ratio of electron-capture to type-II supernovae in single stars
is about $7\%$ if we assume a \citet{1993ApJ...413..641V} wind for
the SAGB phase -- this should be detectable, if it is possible to
distinguish electron-capture from core collapse supernovae.

\section*{Acknowledgements}

\label{sec:Acknowledgements}RGI and AJP are supported by the NWO.

{\footnotesize \bibliographystyle{aa}
\bibliography{izzard-granada2006}
}
\end{document}